\documentclass[pre,twocolumn,showpacs]{revtex4-1}
\usepackage{amsmath}
\usepackage{graphicx}
\usepackage{color}
\usepackage{amssymb}
\usepackage{wasysym}
\usepackage{epstopdf}
\input{epsf}
\usepackage{bm}% bold math
\usepackage{amsmath}
\usepackage{amssymb}

\begin{document}

\title{Dynamics of Hard Colloidal Cuboids in Nematic Liquid Crystals}

\author{Alejandro Cuetos$^1$}
\email{acuemen@upo.es}
\author{Alessandro Patti$^2$}
\email{alessandro.patti@manchester.ac.uk}
\affiliation{$^1$Department of Physical, Chemical and Natural Systems, Pablo de Olavide University, 41013 Sevilla, Spain}
\affiliation{$^2$Department of Chemical Engineering and Analytical Science, The University of Manchester, Manchester, M13 9PL, UK}

\begin{abstract}

We perform Dynamic Monte Carlo simulations to investigate the equilibrium dynamics of hard colloidal cuboids in oblate and prolate nematic liquid crystals. In particular, we characterise the particles' diffusion along the nematic director and perpendicularly to it, and observe a structural relaxation decay that strongly depend on the particle anisotropy. To assess the Gaussianity of their dynamics and eventual occurrence of collective motion, we calculate two- and four-point correlation functions that incorporate the instantaneous values of the diffusion coefficients parallel and perpendicular to the nematic director. Our simulation results highlight the occurrence of Fickian and Gaussian dynamics at short and long times, locate the minimum diffusivity at the self-dual shape, the particle geometry that would preferentially stabilise biaxial nematics, and exclude the existence of dynamically correlated particles.

\end{abstract}


%\keywords{Colloids; Liquid Crystals; Dynamic Monte Carlo; Collective Dynamics.}

\maketitle

%%%%%%%%%%%%%%
%%%%%%%%%%%%%%

\section{Introduction}

Colloids are systems of macromolecules or nanoparticles dispersed through a continuous medium. If the dispersed phase consists of solid particles and the continuous phase is a liquid, like an ink or a paint, then they are referred to as colloidal suspensions. The particles' ability of remaining suspended in the liquid stems from a thermal energy that must overcome the gravitational potential energy, which, by contrast, promotes sedimentation. Colloidal suspensions of anisotropic particles are complex fluids that display a particularly rich phase behaviour, with liquid crystals (LCs) forming between the isotropic and the crystal phase. LCs are states of matter with intermediate features between those of a simple, disordered liquid and those of a crystalline solid. More specifically, they flow like liquids, but their building blocks are able to orient along one or more spatial directions, thus closely reproducing the internal ordering of molecular crystals and their optical properties.

A renewed interest has been recently devoted to colloidal suspensions of anisotropic particles with a biaxial geometry, such as  board-like (cuboids) and bent-core particles, whose orientation in space is fully identified by two independent orthogonal axes. Biaxial particles are, in principle, excellent candidates to form biaxial nematic (N$_\text{B}$) LCs, whose existence, predicted by a generalized Maier-Saupe theory proposed by Freiser exactly half a century ago \cite{maier, freiser}, is still the object of a fervent research interest among experimental, computational and theoretical groups \cite{vandenpol, belli1, peroukidis1, leferink, raton, Yang, mundoor, Dussi2, orlandi, drwenski, Cuetos1, Patti1, Cuetos2, chiappini}. This interest is rooted in the captivating idea of manufacturing biaxial LCDs, by many envisaged as the next-generation fast display technology, but still at an embryonic stage due to the difficulty of unambiguously identifying stable molecular N$_\text{B}$ phases. By contrast, stable colloidal N$_\text{B}$ phases have been observed for the first time approximately ten years ago in systems of polydisperse prolate goethite particles \cite{vandenpol}.

Despite such an intense research activity aiming at fully unveiling the phase and aggregation behaviour of biaxial particles \cite{luckhurst}, much less attention has been devoted to their dynamics. On one hand, this is not surprising as, apart from the especially successful case of the Gay-Berne fluid \cite{gay} and the purely orientational pair potential proposed by Straley \cite{straley}, most of the model particles with a biaxial geometry (\textit{e.g.} cuboids) lack a suitable interaction potential to compute time trajectories and are often modelled as arrays of spherical beads \cite{cinacchi1}. On the other hand, it is not an easy task to predict how and, especially, over what time scales the N$_\text{B}$ phase would respond to external stimuli and hence quantify the advantages of employing them in LCDs, if an insight into their dynamical behaviour is missing. For instance, it is relevant to know how the shape anisotropy of biaxial particles determines their ability of diffusing in a crowded and ordered mesophase, such as a nematic LC; it is also important to know whether these particles are dynamically correlated as this would influence the response time of the whole phase to applied forces. In this context, the Dynamic Monte Carlo (DMC) simulation method plays a very important role as it provides a theoretical framework to mimic the dynamics of colloidal particles of any possible shape without integrating stochastic or deterministic differential equations as done, respectively, in Brownian Dynamics (BD) or Molecular Dynamics (MD) simulations \cite{Patti2, Cuetos3, Corbett}. The DMC method, which shows excellent quantitative agreement with BD simulations and can be between 10 and 20 times faster than BD \cite{Corbett}, is especially useful to mimic the dynamics of hard-core particles, which cannot be straightaway reproduced by MD or BD simulations. Consequently, the dynamics of biaxial particles interacting via a mere steric repulsion can be analysed by DMC simulations. The main goal of the present work is therefore investigating the equilibrium dynamics of prolate and oblate hard board-like particles (HBPs) in nematic LCs and characterise their structural relaxation decay.

This paper is organised as follows. In Section II, we introduce the model and simulation methodology, referring the reader to our previous contributions for specific details on the implementation of the DMC technique. The relevant correlation functions to assess the dynamics are also discussed in this section. In Section III, we discuss the main features of the equilibrium dynamics of cuboids and, finally, in Section IV, we draw our conclusions.

%%%%%%%%%%%%%%
%%%%%%%%%%%%%%

\section {Model and Simulations}

\begin{figure}[h]
\centering
  \includegraphics[width=1.0\columnwidth]{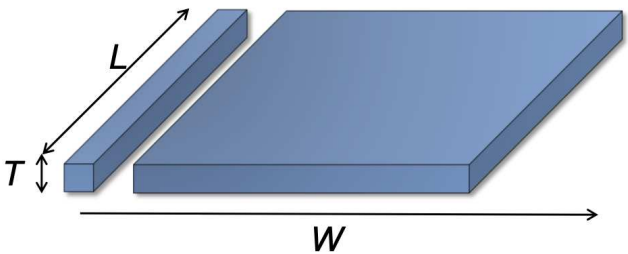}
  \caption{Model HBPs with thickness $T$, length $L=12T$, and width $W$. The particle width is a simulation parameter with values between $W=T$ (rod-like particles) and $W=L$ (disk-like particles).}
  \label{fig1}
\end{figure}

The colloidal cuboids studied in this work are modelled as hard board-like particles (HBPs), whose main geometrical features are shown in Fig.\ \ref{fig1}. They are rectangular prisms of thickness $T$, length $L=12T$ and width $T\leq W \leq L$. The particle thickness, $T$ is our unit length and is kept constant, while the particle width, $W$, is a simulation parameter that varies between $T$ and $L$ in order to reproduce prolate (rod-like) and oblate (disk-like) particles, respectively. At $W=\sqrt{LT}$, HBPs are not oblate nor prolate, but display an ambivalent nature that can produce nematic LCs with oblate, prolate or biaxial symmetry. This special geometry is referred to as self-dual shape. 

The width-to-thickness ratio, $W^* \equiv W/T$, and the system packing fraction, $\eta \equiv N_pv_0/V$, completely determine the phase behaviour and dynamics of these systems. In the latter definition, $V$ is the volume of the simulation box, $N_p=2000$ the number of HBPs, and $v_0=12WT^2$ the particle volume. The phase diagram of monodisperse HBPs, studied by Monte Carlo (MC) simulations, reveals the existence of oblate and prolate nematic and columnar liquid-crystalline phases as well as biaxial smectics \cite{Cuetos1}. More recently, simulations of especially large systems suggested that the prolate columnar phase might actually be metastable with respect to either a smectic phase or a crystal phase, depending on the system density \cite{Dussi1}. Over the last two years, the quest for the elusive biaxial nematic ($\text{N}_\text{B}$) phase has produced a number of theoretical \cite{belli1}, simulation \cite{Dussi2, Patti1, Cuetos2} and experimental \cite{Yang} studies that confirm the difficulty of obtaining a stable $\text{N}_\text{B}$ phase of cuboids. The focus of the present contribution is investigating and characterising the relaxation dynamics of prolate (N$^+$) and oblate (N$^-$) uniaxial nematic phases of HBPs. To this end, we first performed standard MC simulations in the canonical ensemble to equilibrate the systems at the packing fraction, $\eta=0.34$, at which N$^+$ and N$^-$ phases are known to be stable over the full set of aspect ratios studied here, namely  $1 \leq W^* \le \sqrt{L^*} \approx 3.46$ for N$^+$ and $3.46 \le W^* \le 12$ for N$^-$\cite{Cuetos1}. To study the relaxation dynamics of these systems, we performed Dynamic Monte Carlo (DMC) simulations in the canonical ensemble. Details of the DMC method are available to the interested reader in our former works \cite{Patti2, Cuetos3, Corbett}. Here, we briefly summarise the main points that are essential to gain a general insight. 

DMC simulations are performed in cubic boxes with periodic boundaries and, since our intention is to mimic the time evolution of physical systems, no unphysical moves, such as cluster moves or particle swaps, are allowed. One DMC cycle consists of $N_p$ attempts of simultaneously displacing and rotating randomly selected HBPs. These combined moves are accepted or rejected according to the standard Metropolis algorithm, that is with probability $\text{min}[1, \text{exp}(-\beta \Delta E)]$, where $\Delta E$ is the energy difference resulting from the particle's move and $\beta$ the inverse temperature. Because the interactions between HBPs are modelled by a hard-core potential, moves are always accepted unless an overlap is produced. Overlap tests are based on the separating axes algorithm proposed by Gottschalk \textit{et al.} \cite{gottschalk} and later adapted by John and Escobedo to investigate the phase behaviour of colloidal cuboids with square cross section \cite{john}. The position of the center of mass of a given particle $j$ is updated by decoupling the displacement into three contributions: $\delta \textbf{r}_j = X_T \hat{\textbf{u}}_j + X_W \hat{\textbf{v}}_j + X_L \hat{\textbf{w}}_j$, where $\hat{\textbf{u}}_j$, $\hat{\textbf{v}}_j$, and $\hat{\textbf{w}}_j$ are unit vectors parallel to the particle thickness, width and length, respectively. The magnitude of displacements along the three particle axes is selected within uniform distributions that depend on the particle translational diffusivities at infinite dilution, $D^{tra}_{\alpha,j}$, with $\alpha = T$, $W$, or $L$. In particular:

\begin{equation}
    |X_{\alpha}| \le \sqrt{2D^{tra}_{\alpha,j} \delta t_{\text{MC}}},
\end{equation}

\noindent where $\delta t_{\text{MC}}$ is the timescale for a single DMC cycle. As far as the rotations are concerned, we attempt to reorient the particle's axes $\hat{\textbf{w}}_j$, $\hat{\textbf{v}}_j$, and $\hat{\textbf{u}}_j$ by three consecutive rigid rotations around $L$, $W$, and $T$, respectively. These rotations must satisfy the following condition:

\begin{equation}
|Y_{\alpha}| \le \sqrt{2 D^{rot}_{\alpha,j} \delta t_{\text{MC}}},     
\end{equation}

\noindent where $D^{rot}_{\alpha,j}$ are the particle rotational diffusivities at infinite dilution. The diffusion coefficients $D^{tra}_{\alpha}$ and $D^{rot}_{\alpha}$ have been estimated by employing HYDRO$^{++}$, a program that calculates the solution properties of macromolecules and colloidal particles by approximating their shape and volume with an array of spherical beads of arbitrary size \cite{carrasco, delatorre}. Numerical values of the translational and rotational diffusion coefficients used in this work are given in Table I in units of $D_0 \equiv T^2 \tau^{-1}$, where $\tau$ is the time unit. In order to compare the dynamics of the systems listed in Table I, it is crucial to recover the actual timescale of the Brownian dynamics (BD), $\delta t_{\text{BD}}$ \cite{Patti2, Cuetos3, Corbett}. To this end, one needs to rescale the DMC time scale, which has been set to $\delta t_{\text{MC}} = 10^{-2}\tau$ for all the systems studied, via the acceptance rate $\mathcal{A}$:

\begin{equation}
    \delta t_{\text{BD}} = \frac{\mathcal{A}}{3} \delta t_{\text{MC}}.
\end{equation}

\begin{table*}[t]
 \caption{Infinite-dilution translational and rotational diffusion coefficients of HBPs with different width-to-thickness ratios, $W^*$, as obtained from HYDRO$^{++}$ \cite{carrasco, delatorre}.}
\centering
\begin{tabular}{|c||c|c|c||c|c|c|}
\hline
\toprule
$W^*$ & $D^{tra}_T/D_0$ & $D^{tra}_W/D_0$ & $D^{tra}_L/D_0$ & $D^{rot}_T/D_0$ & $D^{rot}_W/D_0$ & $D^{rot}_L/D_0$ \\ \hline
1    & $2.2 \cdot 10^{-2}$ & $2.2 \cdot 10^{-2}$ & $3.1 \cdot 10^{-2}$ & $1.1 \cdot 10^{-3}$ & $1.1\cdot 10^{-3}$ & $2.3 \cdot 10^{-2}$\\ \hline
2.5  & $1.7 \cdot 10^{-2}$ & $1.9 \cdot 10^{-2}$ & $2.4 \cdot 10^{-3}$ & $7.9 \cdot 10^{-4}$ & $7.1 \cdot 10^{-4}$ & $5.8 \cdot 10^{-3}$\\ \hline
3    & $1.6 \cdot 10^{-2}$ & $1.8 \cdot 10^{-2}$ & $2.2 \cdot 10^{-2}$ & $7.3 \cdot 10^{-4}$ & $6.5 \cdot 10^{-4}$ & $4.2 \cdot 10^{-3}$\\ \hline
3.46 & $1.5 \cdot 10^{-2}$ & $1.8 \cdot 10^{-2}$ & $2.1 \cdot 10^{-2}$ & $6.7 \cdot 10^{-4}$ & $6.0 \cdot 10^{-4}$ & $3.2 \cdot 10^{-3}$\\ \hline
4    & $1.4 \cdot 10^{-2}$ & $1.7 \cdot 10^{-2}$ & $2.0 \cdot 10^{-2}$ & $6.2\cdot 10^{-4}$ & $5.6 \cdot 10^{-4}$ & $2.5 \cdot 10^{-3}$\\ \hline
6    & $1.2 \cdot 10^{-2}$ & $1.5 \cdot 10^{-2}$ & $1.7 \cdot 10^{-2}$ & $4.6 \cdot 10^{-4}$ & $4.4 \cdot 10^{-4}$ & $1.1 \cdot 10^{-3}$\\ \hline
8    & $9.4 \cdot 10^{-3}$ & $1.4 \cdot 10^{-2}$ & $1.5 \cdot 10^{-2}$ & $3.5 \cdot 10^{-4}$ & $3.6 \cdot 10^{-4}$ & $6.3 \cdot 10^{-4}$\\ \hline
12   & $6.5 \cdot 10^{-3}$ & $1.2 \cdot 10^{-2}$ & $1.2 \cdot 10^{-2}$ & $2.1 \cdot 10^{-4}$ & $2.6 \cdot 10^{-4}$ & $2.6 \cdot 10^{-4}$\\ \hline
\toprule
\end{tabular}
\end{table*}

\noindent This result has been applied to rescale the dynamical properties that characterise the structural relaxation of our systems, namely (\textit{i}) the mean-square displacement; (\textit{ii}) the self part of the van Hove function; (\textit{iii}) the self-intermediate scattering function; and (\textit{iv}) the dynamic susceptibility.

The mean square displacement (MSD) is the ensemble average of the particles' displacement from their position over a given time window and is defined as

\begin{equation} \label{eq10}
\big \langle \Delta r^{2}(t) \big \rangle= \Bigg \langle \frac{1}{N_p} \sum_{j=1}^{N_p} (\textbf{r}_j(t)-\textbf{r}_j(0))^2 \Bigg \rangle,
\end{equation}

\noindent where $\textbf{r}_j$ is the position vector of particle $j$ and the angular brackets indicate ensemble average over all the HBPs and at least 100 independent trajectories. The self part of the van Hove function (s-VHF) provides the probability distribution of the HBPs' displacements at time $t_0+t$, given their position at time $t_0$ and reads

\begin{equation}
\label{Gtot}
G_s(r,t) =\frac{1}{N_p} \left<\sum_{j=1}^{N_p}\delta(r-|{\bf r}_j(t_0+t)-{\bf r}_j(t_0)|)\right>,
\end{equation}

\noindent where the symbol $\delta$ is the Dirac delta. If the distribution of the displacements over time is Gaussian, then $G_s(r,t)$ can be approximated by a Gaussian function that reads

\begin{equation}\label{eq17}
G'_{s,d}(r,t)=(4\pi D_t t)^{-d/2} \exp{\left(-\frac{r^2}{4D_tt} \right)},
\end{equation}

\noindent  where $D_t$ is the instantaneous diffusion coefficient and $d$ the dimensionality of the system dynamics. In particular, $G'_{s,1}$ and $G'_{s,2}$ are the Gaussian approximations of the parallel and perpendicular s-VHFs. As far as $G'_{s,3}$ is concerned, we have recently shown that it would not correctly determine the Gaussianity of the displacements in anisotropic systems, and thus proposed a Gaussian distribution that has an ellipsoidal, rather than spherical, symmetry \cite{Cuetos4}. More specifically, for prolate particles, the Gaussian approximation of the s-VHF function takes the following form:

\begin{equation}\label{eq17}
G'_{s,3}(r,t)=\frac{\Omega}{(4\pi t)^{3/2}} \exp\left(-\frac{r^2t^{-1}}{4D_{t,\parallel}}\right) \frac{F(r \Delta_p^{1/2})}{r\Delta_p^{1/2}}
\end{equation}

\noindent and for oblate particles

\begin{equation}\label{eq18}
G'_{s,3}(r,t)= \frac{\Omega\sqrt{\pi}}{2(4\pi t)^{3/2}} \exp\left(-\frac{r^2t^{-1}}{4D_{t,\perp}}\right) \frac{\text{erf}(r \Delta_o^{1/2} )}{r \Delta_o^{1/2}},
\end{equation}

\noindent where $D_{t,\parallel}$ and $D_{t,\perp}$ are, respectively, the instantaneous values of the diffusion coefficients in the direction parallel and perpendicular to $\bf{\hat{n}}$, respectively, $F(...)$ is the Dawson's integral, erf(...) the error function, $\Omega=1/(D_{t,\perp}^2D_{t,\parallel})^{1/2}$, and $\Delta_p=-\Delta_o=1/(4D_{t,\perp}t)-1/(4D_{t,\parallel}t)$.

The self-intermediate scattering function (s-ISF) is useful to gain an insight into the timescales associated to the system's structural relaxation occurring over a length $2 \pi / |{\bf k}|$, where $\mathbf{k}$ is a wave vector that corresponds to the location of the relevant peaks in the static structure factor. It is defined as

\begin{equation}
F_s(k,t) = \frac{1}{N_p} \left< \sum_{j=1}^{N_p} \exp[i \mathbf{k} \cdot ( \mathbf{r}_{j}(t+t_{0}) - \mathbf{r}_{j}(t_{0})) ] \right>.
\end{equation}

\noindent The fluctuations of the self-intermediate scattering function define the so-called dynamical susceptibility, $\chi_4 (k,t)$, which provides useful information on the extent of dynamic heterogeneities or, equivalently, on the number of dynamically correlated particles:

\begin{equation}
\chi_4 (k,t) = N_p \big[ \big \langle f^2_s(k,t) \big \rangle - F_s^2({k,t}) \big ]
\end{equation}

\noindent where $f_s(k,t) = \sum_{j=1}^{N_p} \cos({\bf k}[ ( \textbf{r}_j(t+t_0) - \textbf{r}_{j}(t_0))]/N_p$ is the real part of the instantaneous value of the s-ISF.

%%%%%%%%%%%%%%%%%%%%%%%%%%%%%%%%%%%%%%%%%%%%%%%%%%%%%%%%%%%%%
%%%%%%%%%%%%%%%%%%% RESULTS %%%%%%%%%%%%%%%%%%%%%%%%%%%%%%%%%
%%%%%%%%%%%%%%%%%%%%%%%%%%%%%%%%%%%%%%%%%%%%%%%%%%%%%%%%%%%%%

\section {Results}

\begin{figure}[b]
\centering
  \includegraphics[width=1.0\columnwidth]{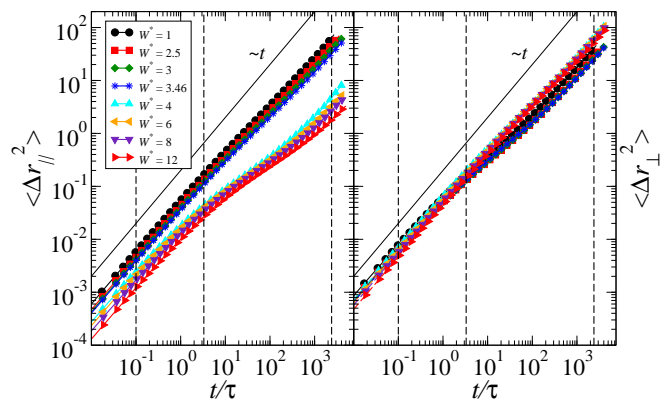}
\caption{Mean square displacement of oblate and prolate HBPs with width-to-thickness ratio indicated in the legend. Left and right frame refer to the directions parallel and perpendicular to the nematic director, respectively. The vertical dashed lines correspond to $t/\tau=0.1$, 3.3 and 2400. The solid lines highlight the linear dependence of the MSD on time at short and long timescales.}
\label{msd}
\end{figure}

In this section, we analyse the main features of the equilibrium dynamics of HBPs in N$^-$ and N$^+$ phases by DMC simulations. All the results shown here have been rescaled with the acceptance rate to recover the time scale of the Brownian dynamics \cite{Patti2, Cuetos3, Corbett}. We stress that this is an essential step in order to correctly compare the dynamic properties of different systems. The typical behaviour detected in dense colloidal LCs generally consists of a short-time diffusive regime, where the MSD depends linearly on time, followed by an intermediate regime where the particles' motion is slowed down by the surrounding neighbours, referred to as caging effect, and a long-time diffusion regime controlled by the collisions between particles \cite{bier, patti3, cinacchi2, matena, patti4, belli2, patti5, piedrahita, peroukidis2, morillo}. The intermediate regime becomes more and more relevant at increasing densities, with the MSD developing a plateau that can extend several time decades in especially dense colloidal suspensions, such as columnar LCs \cite{patti3}. Nematic LCs of uniaxial particles usually show a very short, almost evanescent, intermediate regime, especially at densities very close to the isotropic-nematic phase transition. Similar tendencies are also noticed here in N$^-$ and N$^+$ phases of biaxial HBPs. In particular, the MSD in the direction of $\hat{\bf{n}}$, shown in the left frame of Fig.\ \ref{msd}, reveals a relatively smooth transition from short-time to long-time diffusion in systems of prolate geometries ($1 \le W^* \le 3.46$). Oblate HBPs show a slightly different behaviour that suggests a more pronounced caging effect as compared to that of their prolate counterparts. In this case, the inset of the long-time diffusive regime, where the effect of caging vanishes completely and $\langle \Delta r^2_{\parallel} \rangle$ recovers its linearity with time, occurs between $t/\tau \approx 2 \times 10^3$ and $3 \times 10^3$, depending on the particle's anisotropy. 

\begin{figure}[h]
\centering
  \includegraphics[width=1.0\columnwidth]{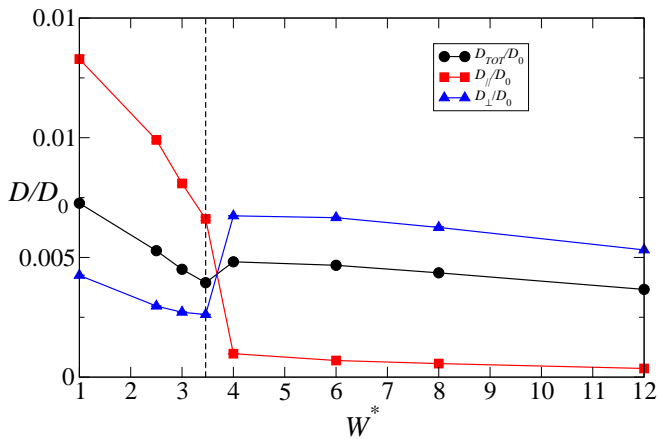}
\caption{Total (\CIRCLE), parallel (\textcolor{red}{$\blacksquare$}) and perpendicular (\textcolor{blue}{$\blacktriangle$}) diffusion coefficients of oblate and prolate HBPs in nematic LCs. The vertical dashed line at $W^*=\sqrt{L^*} \approx 3.46$ indicates the transition from prolate to oblate particle shapes. The solid lines are a guide for the eyes.}
\label{diff_coef}
\end{figure}

By contrast, the perpendicular MSD, $\langle \Delta r^2_{\perp} \rangle$, reported in the right frame of Fig.\ \ref{msd}, is not especially influenced by the particle anisotropy, although it can be observed that, at long time-scales, the MSD of oblate HBPs is slightly larger than that of prolate HBPs. To clarify the role of the shape anisotropy on the particles' ability of diffusing in nematic LCs, we have calculated the diffusion coefficients as $D = \lim_{t\to\infty} \langle \Delta r^2 \rangle/2dt$, where $d=1$, 2 or 3 for parallel, perpendicular or total MSD, respectively. The resulting diffusivities are reported in Fig.\ \ref{diff_coef} as a function of $W^*$. It is interesting to notice that at the prolate-oblate crossover, indicated by the vertical dashed line $W^*=\sqrt{L^*} \approx 3.46$, the parallel diffusivity, $D_{\parallel}$, drops abruptly by a factor of 7, from approximately $6.5 \times 10^{-3}D_0$ at $W^*=3.46$ to $10^{-3}D_0$ at $W^*=4$. An opposite tendency is detected in the perpendicular diffusivity, $D_{\perp}$, which increases from $2.5 \times 10^{-3}D_0$ at $W^*=3.46$ to $7 \times 10^{-3}D_0$ at $W^*=4$. The combined effect of parallel and perpendicular diffusion results in a total diffusion coefficient, $D_{TOT}=(2D_{\perp}+D_{\parallel})/3$ in Fig.\ \ref{diff_coef}, that displays a minimum exactly at the self-dual shape. Despite being the geometry that would more easily stabilise the biaxial nematic phase, the self-dual shape delays diffusion, at least for the range of particle anisotropies explored in this work. 

\begin{figure}[b]
\centering
  \includegraphics[width=1.0\columnwidth]{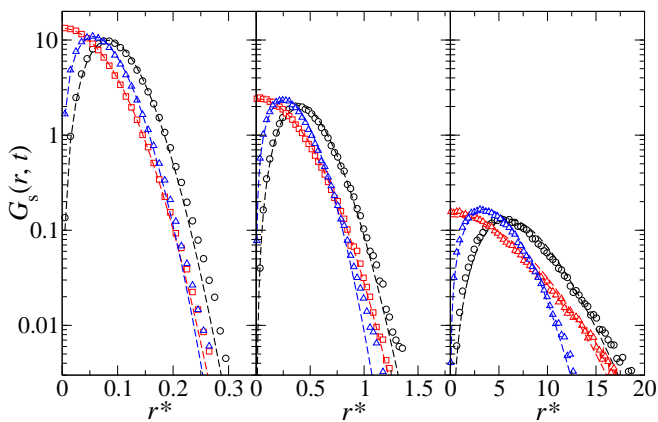}
\caption{Total ($\Circle$), parallel (\textcolor{red}{$\square$}) and perpendicular (\textcolor{blue}{$\triangle$}) self-part of the van Hove correlation function of prolate HBPs with $W^*=1$. Left, middle and right frames refer, respectively, to $t/\tau=0.1$, 3.3 and 2400. Dashed lines are Gaussian distributions obtained from Eqs.\ (6) and (7).}
\label{VHF_w1}
\end{figure}

\begin{figure}[b]
\centering
  \includegraphics[width=1.0\columnwidth]{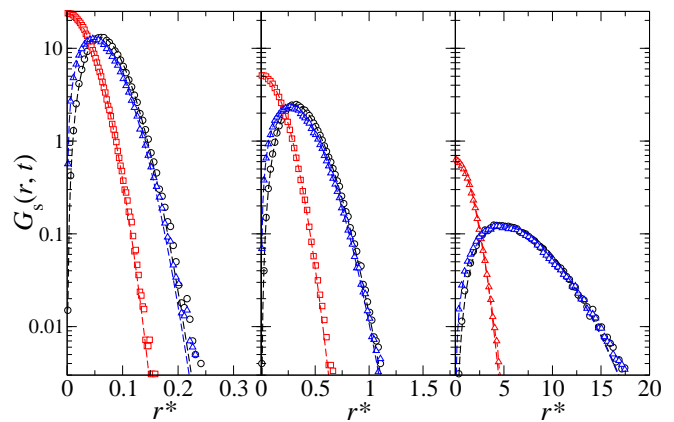}
\caption{Total ($\Circle$), parallel (\textcolor{red}{$\square$}) and perpendicular (\textcolor{blue}{$\triangle$}) self-part of the van Hove correlation function of oblate HBPs with $W^*=12$. Left, middle and right frames refer, respectively, to $t/\tau=0.1$, 3.3 and 2400. Dashed lines are Gaussian distributions obtained from Eqs.\ (6) and (8).}
\label{VHF_w12}
\end{figure}

The effect of shape anisotropy is not limited to the single particle ability to diffuse, but extends to the whole system by contributing to determine the degree of Gaussianity of its dynamics as well as the decay of its structural relaxation. More specifically, eventual deviations from Gaussian dynamics are estimated by computing the self-van Hove correlation functions, while the long-time relaxation of the system's density fluctuations is assessed via the self-intermediate scattering function. The self-van Hove correlation functions for nematic LCs of HBPs with $W^*=1$ and 12 are shown, respectively, in Figs.\ \ref{VHF_w1} and \ref{VHF_w12}. Each frame in the two figures portrays the total (circles), parallel (squares) and perpendicular (triangles) s-VHF at a specific time: $t/\tau=0.1$ (left frame) represents a generic instant within the short-time diffusion; $t/\tau=3.3$ (middle frame) approximately indicates the cross-over from the short-time diffusion to the long-time diffusion, and matches the beginning of the caging effect for the parallel diffusion of oblate HBPs; finally, $t/\tau=2400$ (right frame) is a representative time within the fully-developed long-time diffusive regime.

The Gaussianity of the parallel and perpendicular s-VHFs has been estimated by comparing the simulation results with the Gaussian distribution of Eq.\ (6), with $d=1$ and 2, respectively. By contrast, the Gaussianity of the total s-VHFs has been assessed with Eq.\ (7) (prolate HBPs) and (8) (oblate HBPs). The agreement between DMC simulations and theory is excellent, with almost insignificant deviations from a pure Gaussian distribution of displacements detected in systems of prolate HBPs. These tiny deviations, predominantly observed at long distances, suggest the existence of a number of fast particles that are able to displace longer distances than those normally expected. By contrast, at short distances, no significant deviations between simulation and theory are noticed, thus indicating an unambiguous Gaussian behaviour and the absence of particularly slow particles. The s-VHFs of HBPs with $1<W^*<12$ (not shown here) display a similar Gaussian behaviour, the main differences being the shift of the peak towards larger distances and the gradually lower probability of observing fast particles at increasing $W^*$. It should also be noticed that the essentially Gaussian nature of the s-VHFs excludes, at both short and long times, where the MSD is a linear function of time, the occurrence of Fickian yet non-Gaussian (FNG) dynamics, as we had already observed in nematics of uniaxial particles \cite{Cuetos4}. This finding further confirms that not all soft-matter systems, now also including complex fluids of biaxial particles, necessarily exhibit FNG dynamics.

\begin{figure}[t]
\centering
  \includegraphics[width=1.0\columnwidth]{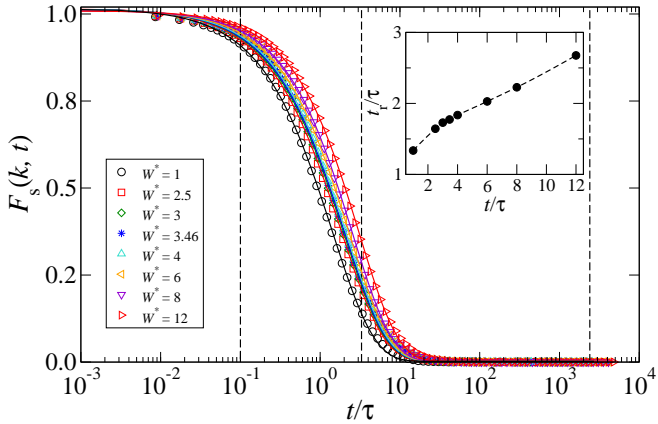}
\caption{Self-intermediate scattering functions calculated at $|{\bf k}|=2\pi/T$ for oblate and prolate HBPs, respectively. The particle width-to-thickness ratio is reported in the legend. The solid lines are stretched exponential fits of the form $\exp{\left[ -(t/t_r)^\alpha) \right]}$, with $t_r$ the relaxation time and $\alpha$ the stretching exponential (see text for details). The inset shows the relaxation time as a function of the particle anisotropy at $|{\bf k}|=2\pi/T$. The solid lines are guides for the eye and the vertical dashed lines correspond to $t/\tau=0.1$, 3.3 and 2400.}
\label{Fs}
\end{figure}

In the light of these considerations, we now discuss the decay of the  density fluctuations in the N$^+$ and N$^-$ phases by examining the self-intermediate scattering functions, $F_s(k,t)$, shown in Fig.\ \ref{Fs}. The s-ISFs of oblate and prolate HBPs have been calculated by randomly selecting $5\times 10^4$ wave vectors ${\bf k}=(2\pi i_1/l, 2\pi i_2/l, 2\pi i_3/l)$, with $i_1$, $i_2$ and $i_3$ integer numbers and $l$ the side of the cubic simulation box, whose magnitude is $|{\bf k}|=2\pi/T$. All curves in Fig.\ \ref{Fs} display a single-step decay, which can be closely reproduced by a stretched exponential function of the type $\exp{\left[ -(t/t_r)^\alpha) \right]}$, with $t_r$ the relaxation time of the system's structural decay and $\alpha$ the stretching exponent, which ranges between 0.82 at $W^*=1$ and 0.88 at $W^*=12$. The former corresponds to the time the s-ISF takes to decays to $e^{-1}$ and it is reported in the inset of Fig.\ \ref{Fs} as a function of the particle reduced width. The latter provides an insight into the nature of the long-time relaxation dynamics. In particular, $\alpha<1$ would indicate the emergence of separated domains of particles with different relaxation dynamics, usually referred to as dynamic heterogeneity. This behaviour, often detected in supercooled liquids, glasses and gels \cite{kob, zheng, colombo}, it is generally attributed to particles that either (\textit{i}) relax exponentially at different time rates or (\textit{ii}) relax non-exponentially at the same time rates \cite{richert, ediger}. 

\begin{figure}[h]
\centering
  \includegraphics[width=1.0\columnwidth]{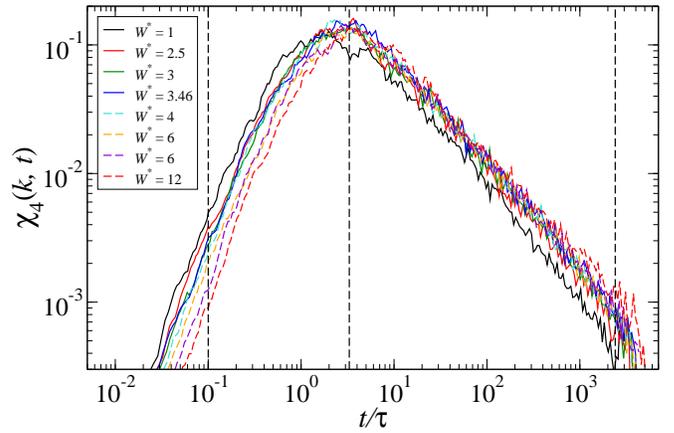}
\caption{Dynamic susceptibility in N$^+$ and N$^-$ LCs of hard cuboids at $|{\bf k}|=2\pi/T$. The particle width-to-thickness ratio is reported in the legend. The vertical dashed lines correspond to $t/\tau=0.1$, 3.3 and 2400.}
\label{chi4}
\end{figure}

To assess the origin of dynamic heterogeneity in nematic LCs of HBPs, we calculated the dynamic susceptibility, $\chi_4(k,t)$, which is proportional to the number of dynamically correlated particles over a time $t$ and a distance $\approx 2\pi/k$. This function hence ponders the presence of clusters that, if existed, would be formed by particles relaxing at the same time rate, thus corroborating scenario (\textit{ii}). The dynamic susceptibility of N$^+$ and N$^-$ phases of HBPs, reported in Fig.\ \ref{chi4}, shows a peak at $t/\tau \approx 3.3$, at the crossover from short-time to long-time diffusive regime. The magnitude of this peak is not significant and, as expected by the relatively large value of the stretching exponent $\alpha$, which is not much lower than 1, excludes the occurrence of HBPs that are dynamically correlated. In other words, in agreement with uniaxial disks and rods \cite{morillo}, there is no evidence of clusters of particles moving collectively and consequently the structural relaxation of both oblate and prolate nematics of HBPs is entirely determined by the single-particle dynamics.

\section{Conclusions}

In summary, we have studied the equilibrium dynamics of prolate and oblate hard cuboids in nematic LCs. To this end, we have employed the DMC simulation technique, which can accurately reproduce the Brownian dynamics of colloids. In particular, we find Fickian and Gaussian dynamics at both short-time and long-time scales, which are separated by a very smooth crossover indicating an almost completely negligible caging effect. The cuboids' anisotropy (precisely, their width-to-thickness ratio) plays a relevant role in the particle's ability to diffuse along the nematic director, $\hat{\bf n}$, and perpendicularly to it. Prolate and oblate cuboids diffuse faster in the direction parallel and perpendicular to $\hat{\bf n}$, respectively. Interestingly, despite being indicated as the most appropriate geometry for the formation of the elusive biaxial nematic phase, the self-dual shape reduces the particles' total diffusivity, which displays a minimum exactly at $W^*=\sqrt{L^*}$. The analysis of the self-van Hove correlation functions reveals the full Gaussianity of the distribution of displacements over time. Very small deviations are actually observed in N$^+$ phases at large distances, suggesting the existence of few fast particles that are not found in N$^-$ phases. Finally, the structural relaxation decay, investigated with the self-intermediate scattering functions, exhibits a stretched exponential behaviour with a relaxation time that approximately increases by a factor 2 with increasing particle width-to-thickness ratio from $W^*=1$ to 12. The stretched exponential nature of the structural relaxation of dense systems, such as supercooled liquids, glasses and gels, usually suggests the occurrence of particles dynamically correlated and eventually displaying collective dynamics \cite{kob, zheng, colombo}. To assess the emergence of this phenomenon, we calculated the four-point dynamic susceptibility, which measures the magnitude of the dynamic correlations. However, as also noticed for uniaxial particles in nematic LCs \cite{morillo}, no sign of dynamic clusters and collective dynamics has been detected. Their occurrence in positionally ordered LC phases is currently under investigation.

\begin{acknowledgements}

AC acknowledges the Spanish Ministerio de Ciencia, Innovaci\'on y Universidades and FEDER for funding (project PGC2018-097151-B-I00) and C3UPO for the HPC facilities provided. AP acknowledges financial support from the Leverhulme Trust Research Project Grant RPG-2018-415.

\end{acknowledgements}

%%%%%%%%%%%%%%%%%%%%%%%%%%%%
%%%%%%%%%%%%%%%%%%%%%%%%%%%%

\end{document}